\documentclass[twocolumn,aps,superscriptaddress,showpacs,nofootinbib,floatfix]{revtex4-1}
\usepackage{graphics}
\usepackage{amsmath}
\usepackage{graphicx}
\usepackage{verbatim}
\usepackage{color}

\begin{document}

\def\Journal#1#2#3#4{{#1} {\bf{#2}}, {#3} (#4).}
\def\ANP{Adv. Nucl. Phys.}
\def\ARNPS{Ann. Rev. Nucl. Part. Sci.}
\def\CTP{Commun. Theor. Phys.}
\def\EPJA{Eur. Phys. J. A}
\def\EPJC{Eur. Phys. J. C}
\def\IJMPA{International Journal of Modern Physics A}
\def\IJMPE{International Journal of Modern Physics E}
\def\JCHP{J. Chem. Phys.}
\def\JCP{Journal of Computational Physics}
\def\JHEP{JHEP}
\def\JPCS{Journal of Physics: Conference Series}
\def\JPG{J. Phys. G: Nucl. Part. Phys.}
\def\NATURE{Nature}
\def\NC{La Rivista del Nuovo Cimento}
\def\NCA{IL Nuovo Cimento A}
\def\NPA{Nucl. Phys. A}
\def\NST{Nuclear Science and Techniques}
\def\PA{Physica A}
\def\PAN{Physics of Atomic Nuclei}
\def\PHY{Physics}
\def\PRA{Phys. Rev. A}
\def\PRC{Phys. Rev. C}
\def\PRD{Phys. Rev. D}
\def\PLA{Phys. Lett. A}
\def\PLB{Phys. Lett. B}
\def\PLD{Phys. Lett. D}
\def\PRL{Phys. Rev. Lett.}
\def\PL{Phys. Lett.}
\def\PREV{Phys. Rev.}
\def\PREP{\em Physics Reports}
\def\PROG{Progress in Particle and Nuclear Physics}
\def\RPP{Rep. Prog. Phys.}
\def\RDNC{Rivista del Nuovo Cimento}
\def\RMP{Rev. Mod. Phys}
\def\SCIENCE{Science}
\def\ZPA{Z. Phys. A.}

\def\ANN{Ann. Rev. Nucl. Part. Sci.}
\def\ANNAST{Ann. Rev. Astron. Astrophys.}
\def\AP{Ann. Phys}
\def\APJ{Astrophysical Journal}
\def\APJS{Astrophys. J. Suppl. Ser.}
\def\EJP{Eur. J. Phys.}
\def\LANC{Lettere Al Nuovo Cimento}
\def\NCA{Nuovo Cimento A}
\def\PHYS{Physica}
\def\NP{Nucl. Phys}
\def\MATH{J. Math. Phys.}
\def\JPAM{J. Phys. A: Math. Gen.}
\def\PRO{Prog. Theor. Phys.}
\def\NPB{Nucl. Phys. B}


\title{Systematic investigation of the particle spectra in Heavy-ion collisions at the Large Hadron Collider}

\author{Hua Zheng}
\affiliation{School of Physics and Information Technology, Shaanxi Normal University, Xi'an 710119, China}
\author{Xiangrong Zhu}
\affiliation{School of Science, Huzhou University, Huzhou 313000, China}
\author{Lilin Zhu}
\affiliation{Department of Physics, Sichuan University, Chengdu 610064, China}
\author{Aldo Bonasera}
\affiliation{Cyclotron Institute, Texas A\&M University, College Station, TX 77843, USA }
\affiliation{Laboratori Nazionali del Sud, INFN, via Santa Sofia, 62, 95123 Catania, Italy}


\begin{abstract}
We investigate the charged particle spectra produced in the heavy-ion collisions at nine centralities from different systems, i.e., Pb+Pb at $\sqrt{s_{NN}}=2.76$ TeV and 5.02 TeV as well as Xe+Xe at $\sqrt{s_{NN}}=5.44$ TeV, at Large Hadron Collider (LHC) using one empirical formula inspired by the solution of the Fokker-Planck equation, dubbed as the generalized Fokker-Planck solution (GFPS). Our results show that the GFPS can reproduce the experimental particle spectrum up to transverse momentum $p_T$ about 45 GeV/c with the maximum discrepancy 30\% covering 10 orders of magnitude. The discrepancy between the data and the results from the GFPS decreases to 15\% when the maximum of the charged particle transverse momentum is cut to 20 GeV/c. We confirmed that the Tsallis distribution derived from the non-extensive statistics, which can reproduce the particle spectra produced in small collision systems, such as p+p, up to few hundreds GeV/c, can only apply to systematically study the particle spectra up to 8 GeV/c in A+A collisions at LHC, as pointed out in the study of identified particle spectra in Pb+Pb collisions at  $\sqrt{s_{NN}}=2.76$ TeV.  The possible explanation why GFPS functions well is also discussed.
\end{abstract}

\pacs{25.75.Nq, 25.75.Ld}
\keywords{}
\maketitle

\section{introduction}
The advent of a new generation of high energy colliders, such as the Relativistic Heavy-Ion Collider (RHIC) at Brookhaven and the Large Hadron Collider (LHC) at CERN, has launched a new era for the researchers to explore the properties of a deconfined quark-gluon plasma (QGP) that could be created in the heavy-ion collisions, including their dynamical evolution, and study the particle production in those collisions. However, the QGP cannot be probed directly experimentally because of the nature of the confinement. Therefore, the particle spectrum, which carries the information of the fireball produced in the collisions and the particle production mechanism, is one of the most important observables in the heavy-ion collisions. Up to now, the particle spectra including the inclusive charged particles and identified particles produced from different collision systems and colliding energies have been widely measured at RHIC and LHC \cite{phenix201405, star2007, daupip2006, alice2, aliceS2012, cms3, cmsdata7000, dauk2013, aliceppb2014, auau62pip2007, Acharya:2018qsh, Acharya:2018eaq, Adam:2015kca}. With the advance of the new technology for the particle detection, particles can be measured with an extremely high transverse momentum $p_T$, i.e., covering few hundreds GeV/c in p+p collisions \cite{cms1, cms2} and around 45 GeV/c in p+A and A+A collisions \cite{aliceppb2014,  Acharya:2018qsh, Acharya:2018eaq, Adam:2015kca}. Thus it becomes more challenging to reproduce the particle spectra fully covering the experimental measured $p_T$ range within one simple framework.

In the past, the theoretical studies of the particle spectrum have been carried out in the framework of hydrodynamical models \cite{Song:2013qma, Zhu:2015dfa, Pang:2018zzo}, transport models \cite{Lin:2004en, Zhu:2018nev, Sa:2011ye}, recombination models \cite{Hwa:2004ng, Hwa:2003ic, Zhu:2013cza, vg, rf} and perturbative QCD (pQCD) \cite{pq1, pq2, jet}, characterized by the regions of transverse momenta of the produced particles where the models are applicable. These methods are independent of each other with great success in explaining the experimental data and reach a high level of sophistication. Besides these, there still exists other phenomenological models proposed to describe the particle production with different assumptions \cite{wongprd, wong2012, wongarxiv2014, cleymans2, beck, azmiJPG2014, Chen:2016lmx, maciej, tsallisbook, khandai, ahep17hua, twocomp, Zheng:2015tua, Zheng:2015gaa, ahep16hua, tsallis}. 

One example is the Tsallis distribution derived from the non-extensive statistics \cite{tsallisbook, beck, tsallis} which has attracted the attention of many theorists and experimentalists and become a topic of great interest in high energy heavy-ion collisions \cite{cmsdata7000,daupip2006, cms2014, star2007, phenix2011, aliceS2012, alice22, cms3, wongprd, wong2012, wongarxiv2014, cleymans2, beck, azmiJPG2014, khandai, ahep17hua, Zheng:2015tua, Zheng:2015gaa, ahep16hua}. It has been very successful in describing the particle spectra measured at RHIC and LHC, especially for the small collision systems, i.e., p+p and p+A, with only three free parameters, and in a wide variety of research fields (see Ref. \cite{tsallisbook} and references therein). In p+p collisions, it is quite impressive to see that the Tsallis distribution can fit the spectra of identified hadrons and charged particles in a large range of $p_T$ up to 200 GeV/c which covers 15 orders of magnitude \cite{wongprd, wong2012, Zheng:2015tua}. It has been nicely demonstrated that the Tsallis distribution can fit almost all the particle spectra measured at RHIC and LHC so far \cite{Zheng:2015tua, Zheng:2015gaa}. However, it can only reproduce part of the observed spectra of identified particles in central Pb+Pb collisions at $\sqrt{s_{NN}}=2.76$ TeV, either in the low or in the high $p_T$ region, as shown in Ref. \cite{Zheng:2015gaa}. 

In order to reproduce the particle spectra in the central Pb+Pb collisions at $\sqrt{s_{NN}}=2.76$ TeV where the nicely single Tsallis distribution fails, the multi-component models, such as the double Tsallis distribution \cite{Rybczynski:2014ura, ahep17hua}, the hydrodynamic extension of a two-component model \cite{twocomp} and the multi-sources model \cite{Chen:2016lmx}, are proposed by considering that the particles at different $p_T$ regions have different origins or produced by different mechanisms. The price to pay for this kind of solution is the increase of the fitting degrees of freedom and sacrifices the beauty and simplicity of the distribution function. Still, it is a natural way out to solve the issue with a clear physical picture. Further, we found that the double Tsallis distribution can reproduce the data for pions and kaons very well, while for protons the data were overestimated at low $p_T$ for central and less central collisions \cite{Rybczynski:2014ura}. 

\begin{figure*}
\centering
\begin{tabular}{ccc}
\includegraphics[width=0.32\textwidth]{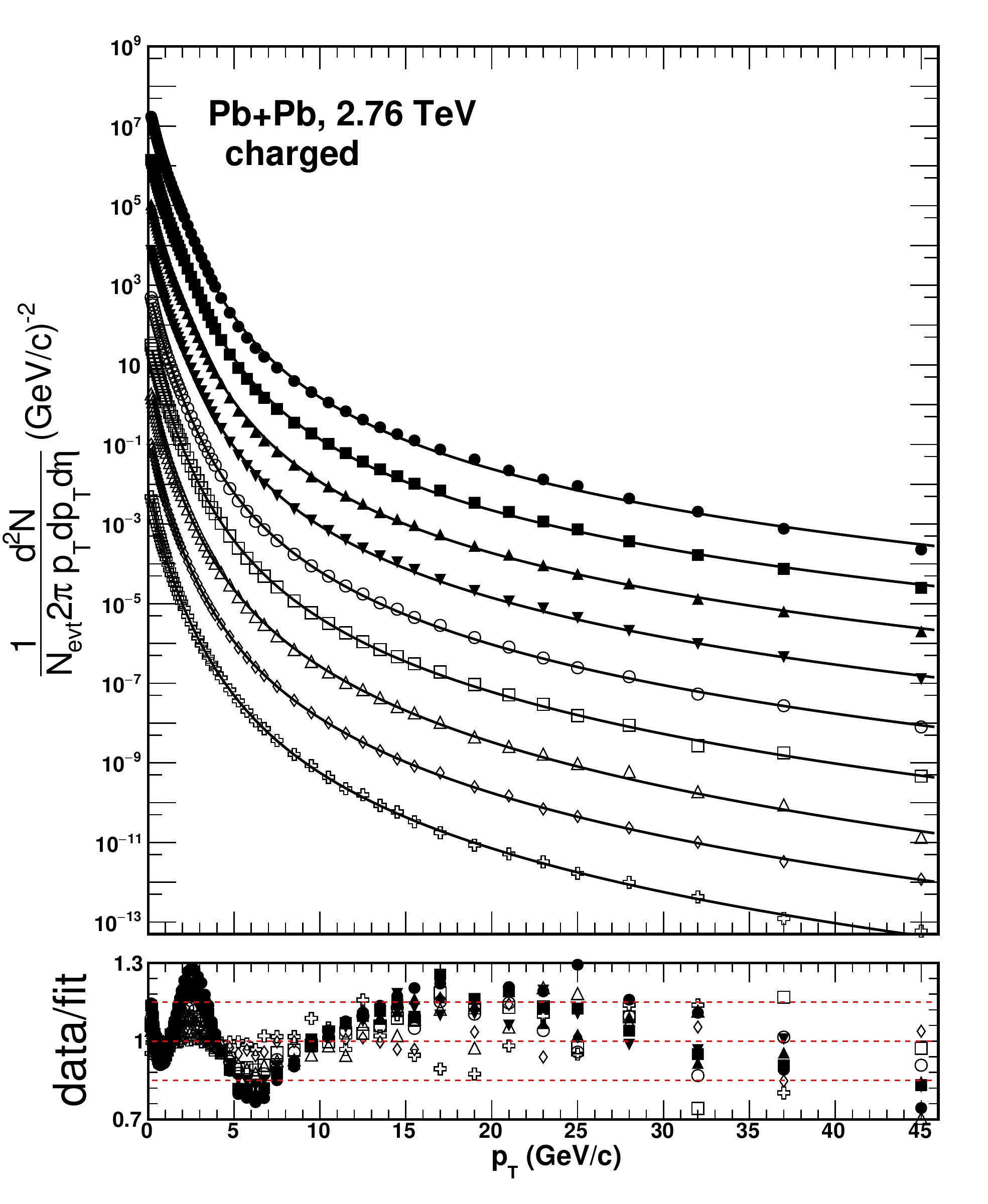}
\includegraphics[width=0.32\textwidth]{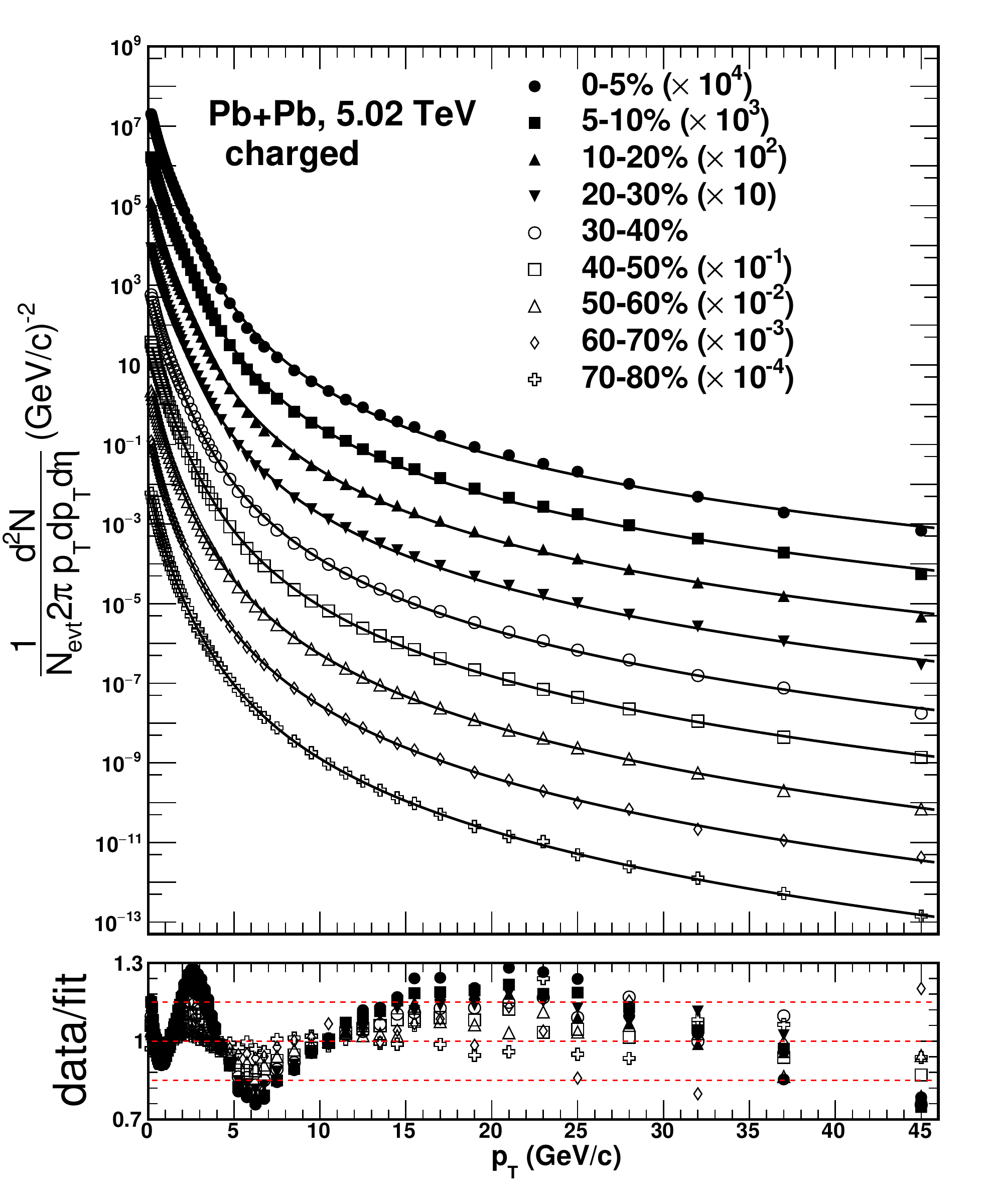}
\includegraphics[width=0.32\textwidth]{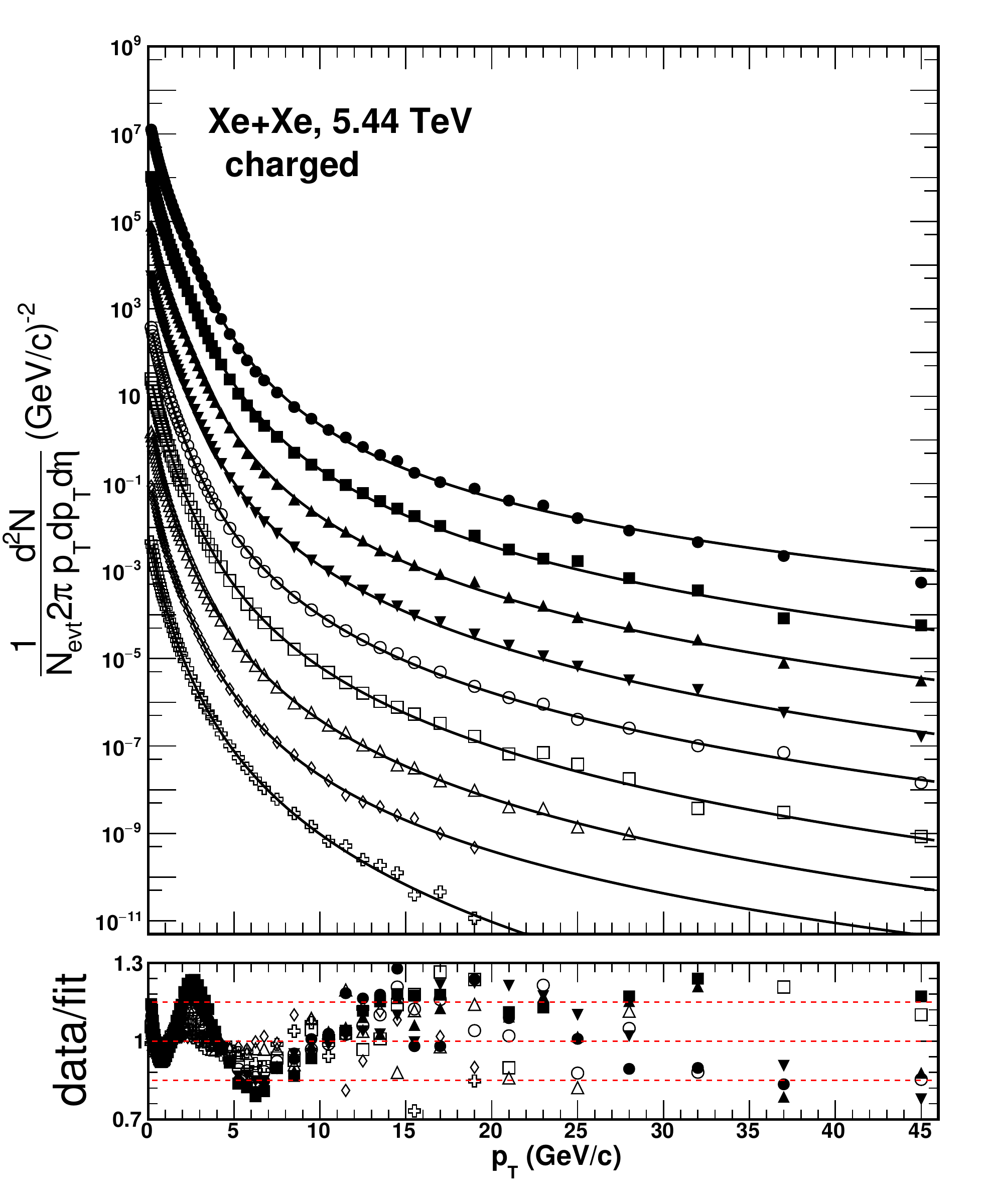}
\end{tabular}
\caption{Fitting results showed by black lines using the generalized Fokker-Planck solution (GFPS) Eq. (\ref{formula2}) for Pb+Pb collisions at $\sqrt{s_{NN}}=2.76$ TeV (left panel) and 5.02 TeV (middle panel), Xe+Xe collisions at $\sqrt{s_{NN}}=5.44$ TeV (right panel) up to $p_T$ around 45 GeV/c. For a better visualization both the data and the analytical curves have been scaled by a constant as indicated. Data are from the ALICE Collaboration~\cite{Acharya:2018qsh, Acharya:2018eaq}.}
\label{fig1}
\end{figure*} 

As a phenomenological study, we are very sensitive to the number of free parameters in our model which is often criticized by others who prefer the sophisticated models or {\it ab initio} calculations. Following the philosophy that we should keep the beauty and simplicity of the distribution function and increase the number of fitting degree of freedom by one each time, similar to the transition from the Boltzmann thermal distribution to the Tsallis distribution when the intermediate $p_T$ data for the particle spectra became available. In Ref. \cite{Zheng:2015gaa}, we proposed a formula inspired by the solution of the Fokker-Planck (FP) equation to fit all the particle spectra available at that time. In Pb+Pb central collisions at $\sqrt{s_{NN}}=2.76$ TeV, we changed the power of $\frac{E_T}{b}$ in the denominator of the stationary solution of FP equation from 2 to 4 in order to increase only one free parameter comparing with the Tsallis distribution. More detailed study in Ref. \cite{ahep17hua} established that the power in the formula should be treated as a free parameter if the spectra of identified particles at different centralities in the Pb+Pb collisions should be reproduced. We dubbed this formula as the generalized Fokker-Planck solution (GFPS).

Recently, the experimental data of charged particle spectra at nine centralities in Pb+Pb collisions at $\sqrt{s_{NN}}=2.76$ TeV and 5.02 TeV, in Xe+Xe collisions at $\sqrt{s_{NN}}=5.44$ TeV with a wide $p_T$ range up to $p_T\sim45$ GeV/c have been released by the ALICE Collaboration at LHC. More systems and more colliding energies with particle spectra in wider $p_T$ range will be the new challenges to all the models. Therefore, our objective in this paper is to investigate whether the GFPS proposed by us can reproduce the new experimental data, which can shed the light on understanding the particle production mechanism at LHC. 

The paper is organized as follows. In the next section, we briefly describe the Tsallis distribution and the generalized Fokker-Planck solution as well as their asymptotic behaviors at low and high $p_T$ respectively. Section~\ref{results} will show the fitting results on charged particle spectra at nine centralities from Pb+Pb collisions at $\sqrt{s_{NN}}=2.76$ TeV and 5.02 TeV as well as Xe+Xe collisions at $\sqrt{s_{NN}}=5.44$ TeV with the GFPS and the Tsallis distributions. A brief discussion for the reason why GFPS performs well to reproduce the particle spectra at LHC is presented in Section \ref{discussion}. Finally, a summary is given in Section~\ref{summary}.

\section{The Tsallis distribution and the generalized Fokker-Planck solution}\label{FP}

\begin{figure*}
\centering
\begin{tabular}{ccc}
\includegraphics[width=0.32\textwidth]{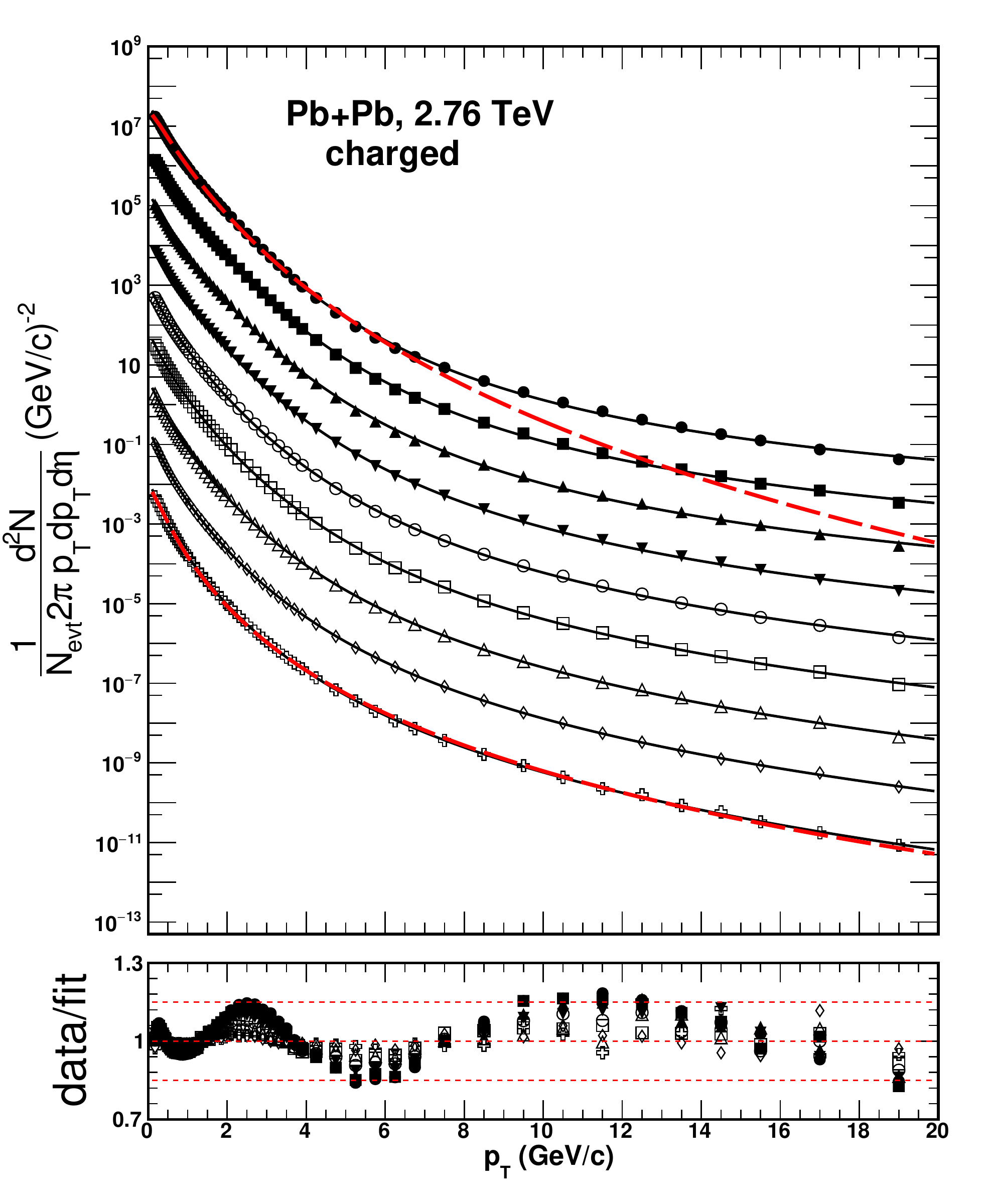}
\includegraphics[width=0.32\textwidth]{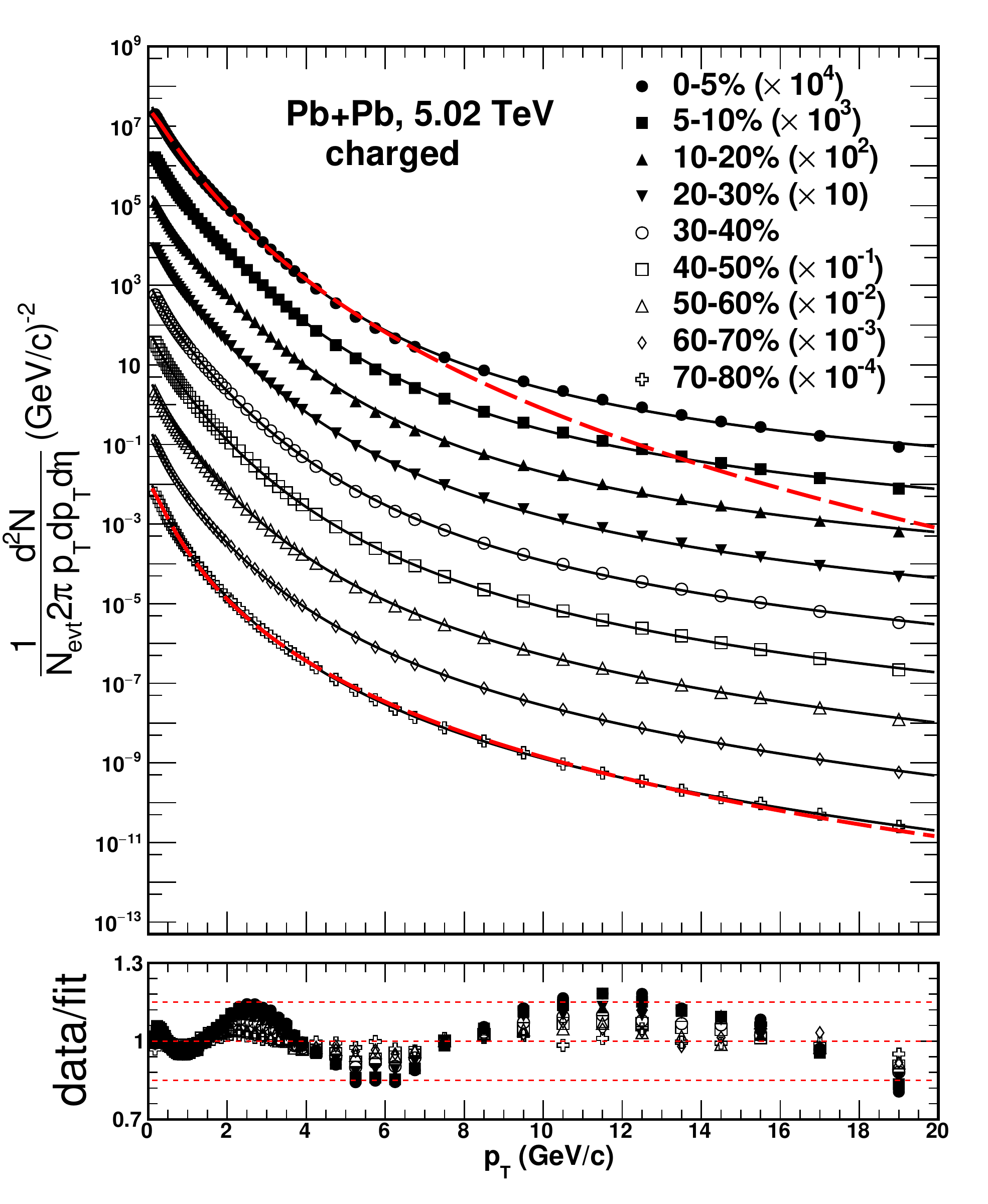}
\includegraphics[width=0.32\textwidth]{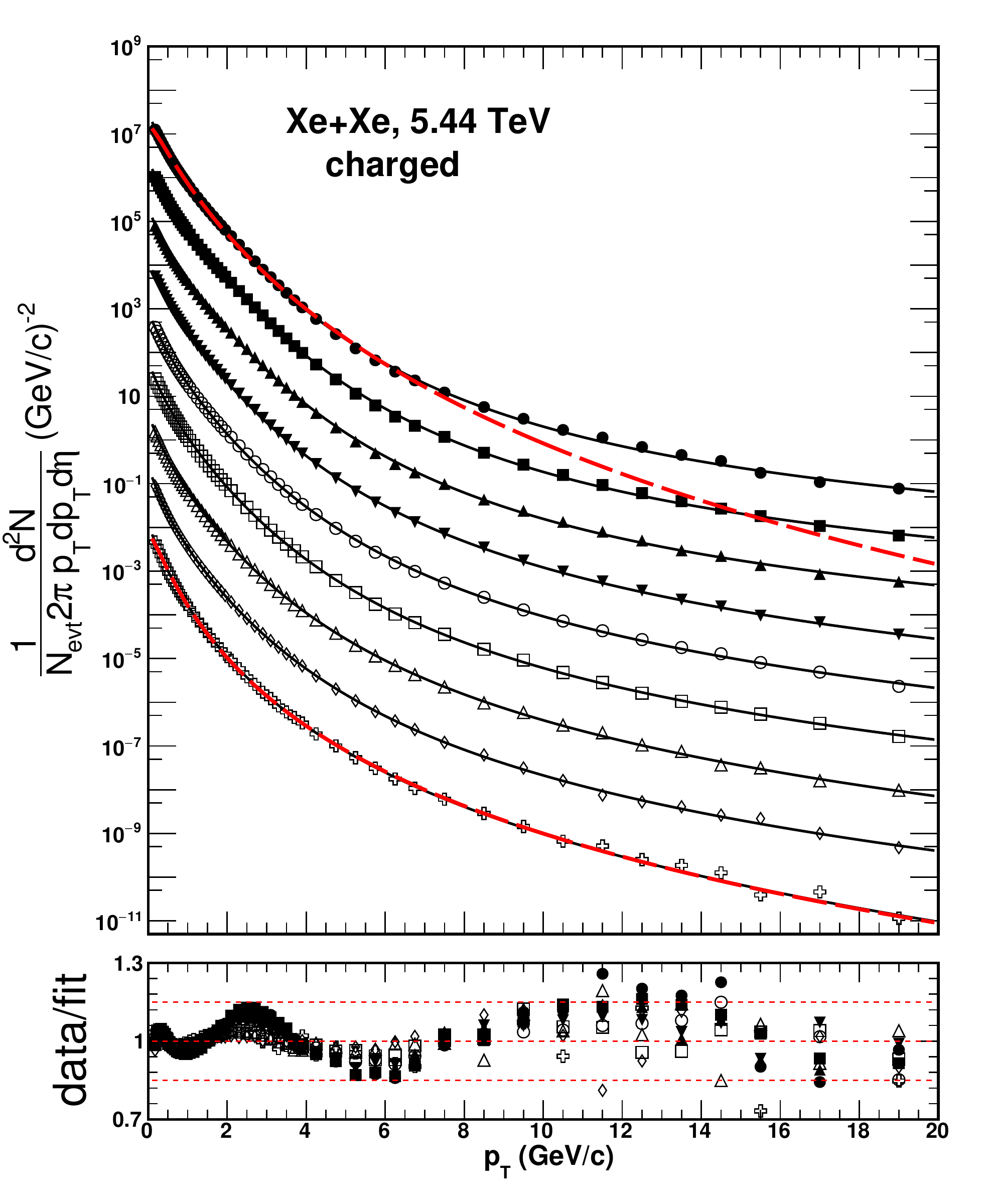}
\end{tabular}
\caption{Fitting results showed by black lines using the generalized Fokker-Planck solution (GFPS) Eq. (\ref{formula2}) for Pb+Pb collisions at $\sqrt{s_{NN}}=2.76$ TeV (left panel) and 5.02 TeV (middle panel) , Xe+Xe collisions at $\sqrt{s_{NN}}=5.44$ TeV (right panel) with $p_T$ cut to 20 GeV/c. The red dashed lines are the results from Tsallis distribution Eq. (\ref{tsallisus}). For a better visualization both the data and the analytical curves have been scaled by a constant as indicated.  Data  are from the ALICE Collaboration~\cite{Acharya:2018qsh, Acharya:2018eaq}.}
\label{fig2}
\end{figure*}

The Tsallis distribution, which was proposed about three decades ago, was derived in the framework of non-extensive statistics \cite{tsallis, tsallisbook}. But when it was applied to the particle spectrum in high energy heavy-ion collisions, different versions of Tsallis distribution appeared in the literature because different considerations were taken into account. In our previous works \cite{Zheng:2015tua, ahep17hua}, we classified those Tsallis distributions into three categories: Type-A, B and C based on their form in order to clarify our considerations and give a nice guide to the researchers interested. We will only briefly introduce the Type-A Tsallis distribution which was adopted by experimental groups.

Type-A Tsallis distribution, which was obtained without resorting to thermodynamical description, has been widely adopted by STAR \cite{star2007}, PHENIX \cite{phenix2011} Collaborations at RHIC and ALICE \cite{alice2, aliceS2012, alice22}, CMS \cite{cms3} Collaborations at LHC. It has the form
\begin{equation}
E\frac{d^3N}{dp^3} = \frac{dN}{dy} \frac{(n-1)(n-2)}{2\pi nC[nC+m(n-2)]}(1+\frac{m_T-m}{nC})^{-n}, \label{exptsallis}
\end{equation}
where $m_T=\sqrt{p_T^2+m^2}$ is the transverse mass. $m$ was used as a fitting parameter in Ref. \cite{star2007}, but it represents the rest mass of the particle studied in Refs. \cite{phenix2011, alice22, alice2, aliceS2012, cms3}.  $\frac{dN}{dy}$, $n$ and $C$ are fitting parameters. The asymptotic behaviors of Eq. (\ref{exptsallis}) can carry out by considering the two limits: 1) When $p_T\gg m$, the $m$ in the last term in Eq. (\ref{exptsallis}) can be neglected and we obtain $E\frac{d^3N}{dp^3} \propto p_T^{-n}$. The particle spectrum follows a power law distribution with $p_T$ is well known in high energy physics; 2) When $p_T\ll m$, i.e., the non-relativistic limit, $m_T -m=\frac{p_T^2}{2m}=E_{T}^{classical}$ and $E\frac{d^3N}{dp^3} \propto e^{-\frac{E_T^{classical}}{C}}$, i.e., a thermal distribution is recovered. The parameter $C$ in Eq. (\ref{exptsallis}) plays the same role as temperature $T$. In Refs. \cite{wong2012, Zheng:2015tua, Zheng:2015gaa}, a simpler form of Eq. (\ref{exptsallis}) is obtained
\begin{equation}
(E\frac{d^3N}{dp^3})_{|\eta|<a} = A(1+\frac{E_T}{nT})^{-n}, \label{tsallisus}
\end{equation}
where $E_T=m_T - m$. $A$, $n$ and $T$ are free fitting parameters in Eq. (\ref{tsallisus}) which is applied in this study. The parameter $n$ determined by the behavior of the particle spectrum at high $p_T$ can be associated with the particle production process as discussed in Ref. \cite{wongprd}.

\begin{figure*}
\centering
\includegraphics[width=0.65\textwidth]{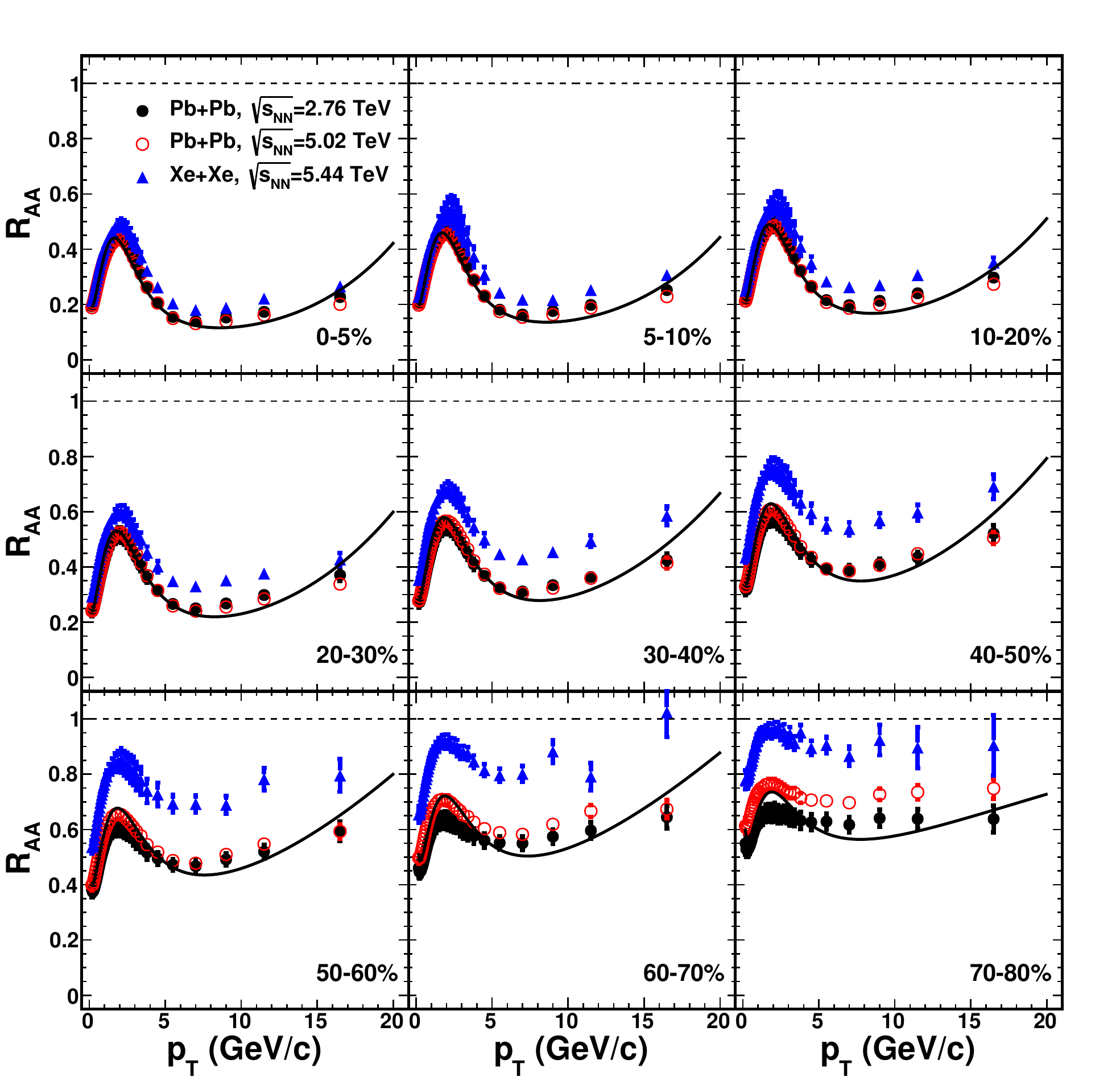}
\caption{The transverse momentum dependence of the nuclear modification factor for charged particles in Pb+Pb collisions at $\sqrt{s_{NN}}=2.76$ TeV (full black circles) and 5.02 TeV (empty red circles), in Xe+Xe collisions at $\sqrt{s_{NN}}=5.44$ TeV (full blue triangles) for nine centrality classes from the ALICE Collaboration~\cite{Acharya:2018qsh, Acharya:2018eaq}. The black lines are the results calculated from the generalized Fokker-Planck solution (GPFS) Eq. (\ref{formula2}) for Pb+Pb collisions at $\sqrt{s_{NN}}=2.76$ TeV.}
\label{fig3}
\end{figure*}

The general form of Fokker-Planck equation is \cite{Banerjee:2010ypa, Svetitsky:1987gq},
\begin{eqnarray}
\frac{\partial P(r,t)}{\partial t}=\frac{\partial }{\partial r}\left[A(r)P(r,t) \right]+\frac{\partial^2 }{\partial r^2}\left[B(r)P(r,t)\right].
\label{formula7} 
\end{eqnarray}
The coefficients $A(r)$ and $B(r)$ are the drift and diffusion terms, respectively.  Here, $r$ represents the variable interested and $t$ is the time. The stationary solution of Eq. (\ref{formula7}) is
\begin{equation}
P_s(r)\propto \frac{1}{B(r)}\exp[-\int^r\frac{A(r')}{B(r')}dr'],
\end{equation}
which fulfills the condition $\frac{\partial P_s}{\partial t}=0$ and depends on the diffusion type through the terms $A(r)$ and $B(r)$. Considering the mixing diffusion $A(E_T)=A_0+\alpha E_T$ and $B(E_T)=B_0+\beta E_T^2$, one could obtain
\begin{equation}
P_s(E_T)=A\frac{e^{-\frac{b}{T}\arctan{\frac{E_T}{b}}}}{[1+(\frac{E_T}{b})^2]^c},
\label{formula8}
\end{equation}
where $b=\sqrt{B_0/\beta}$, $T=B_0/A_0$ and $c=1+\alpha/{2\beta}$. Let us look at the asymptotic behaviors of Eq. (\ref{formula8}): 1) When $p_T\gg 1$ or $\frac{E_T}{b}\gg1$, Eq. (\ref{formula8}) becomes
\begin{eqnarray}
P_s(E_T)\propto p_T^{-2c}.
\label{formula10}
\end{eqnarray}
2) When $p_T\ll 1$ or $\frac{E_T}{b}\ll1$, we obtain
\begin{equation}
P_s(E_T)\propto e^{-\frac{E_T}{T}}.
\label{formula9}
\end{equation}
As one can see that the asymptotic behaviors of $P_s(E_T)$ are consistent with those of Tsallis distribution as well as the particle spectrum distribution in heavy-ion collisions, which exhibits for large $p_T$ roughly a power-law distribution, whereas it becomes purely exponential for small $p_T$. We applied the Eq. (\ref{formula8}) to study the particle spectrum for the first time in central Pb+Pb collisions at $\sqrt{s_{NN}}=2.76$ TeV in Ref. \cite{Zheng:2015gaa} but changed the power of $\frac{E_T}{b}$ in the denominator of the stationary solution of FP equation from 2 to 4. In a more detailed study \cite{ahep17hua}, we realize that we need to generalize the formula proposed in Ref. \cite{Zheng:2015gaa} in order to describe the spectra of identified particles at both central and non-central collisions. Then we adopt the generalized Fokker-Planck solution (GFPS) which is
\begin{eqnarray}
(E\frac{d^3N}{dp^3})_{|\eta|<a} = A\frac{e^{-\frac{b}{T}\arctan{\frac{E_T}{b}}}}{[1+(\frac{E_T}{b})^d]^c}.
\label{formula2}
\end{eqnarray}
It has five parameters $A$, $b$, $c$, $d$ and the effective temperature T (or $\frac{b}{T}$). A crossover from the exponential law at low $p_T$ to the power law at high $p_T$ takes place at $E_T\sim b$.

\begin{table*}[t]
   \caption{The fitting parameters and the corresponding $\chi^2$/ndf for the charged particle spectra of the nine centralities in Pb+Pb collisions at $\sqrt{s_{NN}} = 2.76$ TeV and 5.02 TeV as well as Xe+Xe collisions at $\sqrt{s_{NN}}=5.44$ TeV with Eq. (\ref{formula2}) in Fig. \ref{fig1}.} \label{table}
  \centering
  \begin{tabular}{*{8}{c}}
    \hline
    System  & Centrality & A & $\frac{b}{T}$ & b & c & d &  $\chi^2$/ndf \\
    \hline
     & 0-5\% & 2127.17 & 8.245 & 3.859 & 5.433 & 0.913 & 27.11/56\\
     & 5-10\% &  1822.30 & 8.329 & 4.043 & 5.738 & 0.855 & 25.16/56 \\
     & 10-20\% & 1438.21 & 7.973 & 4.067 & 6.159 & 0.827 & 22.97/56 \\
   Pb+Pb  & 20-30\% &  1030.79 & 7.292 & 3.984 & 6.736 & 0.811 & 17.20/56 \\
    $\sqrt{s_{NN}}=2.76$ TeV & 30-40\% & 748.92 & 6.724 & 4.004 & 7.516 & 0.778 & 11.56/56 \\
    & 40-50\% & 513.68 & 5.996 & 3.861 & 8.164 & 0.766 & 7.33/56\\
    & 50-60\% & 323.20 & 4.868 & 3.549 & 8.878 & 0.774 & 4.82/56 \\
   & 60-70\% &  214.43 & 5.145 & 3.876 & 9.727 & 0.701 & 1.45/56\\
   & 70-80\% & 129.22 & 5.088 & 4.178 & 10.886 & 0.648 & 1.97/56 \\
    \hline
	& 0-5\% &  2453.01 & 8.284 & 4.032 & 5.284 & 0.879 & 36.34/56 \\
	& 5-10\% &  2067.87 & 7.972 & 4.059 & 5.646 & 0.853 & 31.43/56 \\
	& 10-20\% &  1636.11 & 7.693 & 4.106 & 6.032 & 0.822 & 32.10/56 \\
	 Pb+Pb & 20-30\% &  1184.99 & 7.009 & 4.017 & 6.616 & 0.806 & 28.52/56 \\
	$\sqrt{s_{NN}}=5.02$ TeV  & 30-40\% &  851.22 & 6.342 & 3.941 & 7.273 & 0.786 & 24.74/56 \\
	& 40-50\% &  612.44 & 6.068 & 3.980 & 7.869 & 0.744 & 14.38/56 \\
	& 50-60\% &  399.55 & 5.347 & 3.868 & 8.623 & 0.731 & 9.04/56 \\
	& 60-70\% &  262.61 & 5.062 & 3.973 & 9.541 & 0.686 & 3.20/56 \\
	& 70-80\% &  149.93 & 4.783 & 4.056 & 10.365 & 0.654 & 3.50/56 \\
    \hline
	& 0-5\% &  1867.27 & 8.921 & 4.541 & 5.731 & 0.716 & 13.16/55 \\
	& 5-10\% &  1400.69 & 7.462 & 4.246 & 6.481 & 0.794 & 16.98/57 \\
	& 10-20\% &  1124.10 & 6.908 & 4.152 & 6.944 & 0.784 & 16.28/56 \\
	Xe+Xe & 20-30\% &  817.64 & 6.058 & 3.902 & 7.459 & 0.790 & 14.00/56 \\
	$\sqrt{s_{NN}}=5.44$ TeV & 30-40\% &  611.57 & 5.955 & 4.002 & 7.907 & 0.748 & 10.75/56 \\
	& 40-50\% &  424.47 & 5.146 & 3.826 & 8.573 & 0.750 & 7.31/56 \\
	& 50-60\% &  321.27 & 5.333 & 4.137 & 9.355 & 0.682 & 1.45/53 \\
	& 60-70\% &  230.66 & 5.859 & 4.380 & 9.693 & 0.617 & 1.40/49 \\
	& 70-80\% &  100.91 & 2.821 & 3.463 & 11.046 & 0.732 & 0.55/49 \\
    \hline
    \end{tabular}
\end{table*}

\section{results}\label{results}
In the present section, we show the transverse momentum spectra of charged particles in Pb+Pb collisions at $\sqrt{s_{NN}}=2.76$ TeV and 5.02 TeV as well as Xe+Xe collisions at $\sqrt{s_{NN}}=5.44$ TeV and the corresponding fitting results by the generalized Fokker-Planck solution (GPFS) respectively. We also show the nuclear modification factor $R_{AA}$ of the charged particles for the three collision systems and the one calculated from the fitting results for the Pb+Pb collisions at $\sqrt{s_{NN}}=2.76$ TeV.     

\subsection{Transverse momentum spectra}

Figure \ref{fig1} shows the transverse momentum spectra of charged particles in Pb+Pb collisions at $\sqrt{s_{NN}}=2.76$ TeV and 5.02 TeV as well as Xe+Xe collisions at $\sqrt{s_{NN}}=5.44$ TeV. The symbols are the experimental data released by the ALICE Collaboration \cite{Acharya:2018qsh, Acharya:2018eaq} and the black lines are the fitting results from the generalized Fokker-Planck solution, Eq. (\ref{formula2}). The fit metric used is defined by
\begin{eqnarray}
M^2=\sum\limits_i\left[1-\frac{y_i(\textrm{fit})}{y_i(\textrm{data})}\right]^2.
\end{eqnarray}
It is nice to see that all the charged particle spectra from central to peripheral collisions with $p_T$ up to 45 GeV/c covering 10 orders of magnitude are well fitted by the GFPS. Since the data are plotted in log scale, for a better visualization, we also plot the ratios of the data over the fitting results at the bottom in Fig. \ref{fig1}. The ratios clearly establish that the fit to the central collisions has larger deviation from the data than the one to the peripheral collisions which is not surprising as in Ref. \cite{Zheng:2015gaa}, but the maximum deviation is 30\% for all centralities over the full $p_T$ range. The ratios oscillate around 1 indicating that the fitting results with the GFPS are quite reasonable. The fitting parameters in the GFPS are listed in Table \ref{table}, including the information on their chi-squared test. As one can notice that the power index $d$ of $\frac{E_T}{b}$ is less than 1 for all centralities in the three colliding systems, which is smaller than the 2 for the stationary solution in Eq. (\ref{formula8}). We have concluded in Ref. \cite{ahep17hua} that the five parameters in the GFPS are correlated.

Comparing with our previous results \cite{ahep17hua, Zheng:2015gaa}, we notice that the fitting results in Fig. \ref{fig1} are not as good as the previous ones but the fitting $p_T$ range goes up to 45 GeV/c while it is only up to 20 GeV/c in the previous studies. It is well known that the quality of the fitting is strongly dependent on the fitting $p_T$ range. Therefore, in Fig.~\ref{fig2}, we repeat the fitting processes for the charged particle spectra as in Fig. \ref{fig1} but with the $p_T$ cut at 20 GeV/c. Now the fitting results are better than the ones in Fig. \ref{fig1} and the maximum deviation is 15\% for all centralities. In Fig. \ref{fig2}, we also show the fits from Tsallis distribution Eq. (\ref{tsallisus}) with (red) long-dashed lines for the most central collisions (0-5\%) and peripheral collisions (70-80\%). As one can see, the Tsallis distribution can fit the particle spectra at peripheral collisions in AA very well because it is quite similar to the pp collisions and it is not distinguishable from the one with the GFPS. For the central collisions in AA, it can only fit the particle spectra at $p_T<8$ GeV/c because of the strong medium effects. Therefore, it is not the optimal choice to do particle spectra study with full $p_T$ region measured in AA collisions. 

\subsection{The nuclear modification factor}

The nuclear modification factor $R_{AA}$, which is obtained from the ratio of the hadron yield in AA and pp collisions, where the latter is scaled by the average number of binary collisions $\langle N_{coll}\rangle$ in AA collisions,
\begin{eqnarray}
R_{AA}=\frac{1}{\langle N_{coll}\rangle}\frac{d^2N_{AA}/dydp_T}{d^2N_{pp}/dydp_T},
\label{RAA}
\end{eqnarray}
was proposed to show the medium effects in pA or AA collisions. Figure \ref{fig3} shows $R_{AA}$ of charged particles for nine centrality classes in Pb+Pb collisions at $\sqrt{s_{NN}}=2.76$ TeV (full black circles) and 5.02 TeV (empty red circles) and in Xe+Xe collisions at $\sqrt{s_{NN}}=5.44$ TeV (full blue triangles). In the three collision systems, a similar characteristic $p_T$ dependence of $R_{AA}$ is observed, which exhibits a strong centrality dependence with a minimum around $p_T=6-7$ GeV/c (very close to the fitting limit of Tsallis distribution) and an almost linear rise above. In particular, in the most central collisions (0-5\%), the three systems almost show the same values of $R_{AA}$ for the whole $p_T$ region within the errors. While in the peripheral collisions, the suppression of high momentum particles in Pb+Pb collisions is apparently stronger than that in Xe+Xe collisions for the same centrality, which illustrates that the medium is less dense produced in Xe+Xe collisions.

In order to further test the fitting performance of the GFPS to the particles spectra shown in Figs. \ref{fig1} and \ref{fig2}, we calculate the $R_{AA}$ for nine centralities in Pb +Pb collisions at $\sqrt{s_{NN}}=2.76$ TeV using the fitting results in Fig. \ref{fig2}, shown by black lines in Fig. \ref{fig3}. $R_{AA}$ is plotted in linear scale thus very sensitive to the fitting error. As shown in Fig. \ref{fig3}, the calculated results can rather well reproduce the experimental $R_{AA}$,  which is consistent with the results in Figs. \ref{fig1} and \ref{fig2}. Similar results are obtained for the Pb+Pb collisions at $\sqrt{s_{NN}}=5.02$ TeV and Xe+Xe collisions at $\sqrt{s_{NN}}=5.44$ TeV. 

From the results presented in Figs. \ref{fig1}-\ref{fig3}, we can conclude that the generalized Fokker-Planck solution can be a candidate selected to describe the particle spectra in AA collisions at different colliding energies, in particular when the particle spectrum $p_T$ goes higher than 10 GeV/c. 

\section{Discussion}\label{discussion}
In the last section, we just showed that the GFPS can be used in the particle spectrum study phenomenologically to a very high $p_T$ range in the AA collisions at LHC where the Tsallis distribution fails. Actually, the Fokker-Planck equation has been widely employed to study the particle production in heavy-ion collisions \cite{Alberico:2009epja, simon18} and achieved a great success. Therefore, we would like to put our thoughts here why it works well with the lessons learned from the other models. On the one hand, the studies with a multiphase transport (AMPT) model show that the rescatterings among hadrons in the hadronic phase are necessary and essential in order to significantly improve the reproduction of the experimental data \cite{Lin:2004en} and the studies with hydrodynamics model have constrained the QGP shear viscosity to entropy density ratio $\eta/s$ to the range of [0.08, 0.2] which means that the partons in the QGP are strongly correlated similar to the final hadrons \cite{etas1, etas2}. As one can see the common feature of the final hadrons produced in the heavy-ion collisions in the two different frameworks is the particle correlation. On the other hand, in Fokker-Planck equation, there are drift and diffusion terms which reflect the correlations among the particles in the system. We believe that the final particle correlation is the solution why GFPS works to describe the particle spectra produced in AA collisions at LHC according to our study.           

\section{summary}\label{summary}

In this paper, we conducted a detailed study of the transverse momentum spectra of charged particles at nine centralities produced in Pb+Pb collisions at $\sqrt{s_{NN}}=2.76$ TeV and 5.02 TeV as well as Xe+Xe collisions at $\sqrt{s_{NN}}=5.44$ TeV at LHC using the generalized Fokker-Planck solution (GFPS) and the Tsallis distribution. The ratio between the data and fitting result as well as the nuclear modification factor $R_{AA}$ have been adopted to testify the fitting performance of the GFPS. Our results show that the GFPS can nicely describe the charged particle spectra from central to peripheral collisions with $p_T$ up to 45 GeV/c, while the Tsallis distribution can only fit the spectra at $p_T<8$ GeV/c for central collisions and it has the same performance as the GFPS for the peripheral collisions,  which confirms the conclusion made in our earlier work only with the data from Pb+Pb collisions at $\sqrt{s_{NN}}=2.76$ TeV. We also briefly discuss the reason why the GFPS functions well for the particle spectrum.

\section*{Acknowledgements}
This work was supported in part by the NSFC of China under Grant no. 11205106 and The Fundamental Research Funds for the Central Universities (No. GK201903022). The work of Xiangrong Zhu is supported by Zhejiang Provincial Natural Science Foundation of China No. LY19A050001.

\end{document}